\documentclass[journal]{IEEEtran}

\ifCLASSINFOpdf
\else
 \usepackage[dvips]{graphicx}
\fi
\usepackage{url}

\usepackage{graphicx}

\usepackage{amssymb,float,url,times,booktabs, tabularx,color,xcolor,multirow,bbding}
\usepackage[export]{adjustbox} 

\usepackage{amsmath,amsfonts}
\usepackage{algorithmic}
\usepackage{array}
\usepackage[caption=false,font=normalsize,labelfont=sf,textfont=sf]{subfig}
\usepackage{textcomp}
\usepackage{stfloats}
\usepackage{url}
\usepackage{verbatim}
\def\BibTeX{{\rm B\kern-.05em{\sc i\kern-.025em b}\kern-.08em
		T\kern-.1667em\lower.7ex\hbox{E}\kern-.125emX}}
\usepackage{balance}

\usepackage[T1]{fontenc}
\usepackage[OT1]{fontenc}
\usepackage{soul}
\usepackage{subfig}

\begin{document}

\title{Effective Modeling of Critical Contextual Information for TDNN-based Speaker Verification}


\author{Shilong Weng, Liu Yang, Ji Mao
\thanks{S. Weng, L. Yang, and J. Mao are with the School of Computer Science and Cyber Engineering, Guangzhou University, Guangzhou 510006, China (e-mail: shilongweng2010@gmail.com, yanlow2013@gzhu.edu.cn, 2112206046@e.gzhu.edu.cn). Corresponding Author: Liu Yang.}
}


\maketitle

\begin{abstract}
Today, Time Delay Neural Network (TDNN) has become the mainstream architecture for speaker verification task, in which the ECAPA-TDNN is one of the state-of-the-art models. The current works that focus on improving TDNN primarily address the limitations of TDNN in modeling global information and bridge the gap between TDNN and 2-Dimensional convolutions. However, the hierarchical convolutional structure in the SE-Res2Block proposed by ECAPA-TDNN cannot make full use of the contextual information, resulting in the weak ability of ECAPA-TDNN to model effective context dependencies.
To this end, three improved architectures based on ECAPA-TDNN are proposed to fully and effectively extract multi-scale features with context dependence and then aggregate these features.
The experimental results on VoxCeleb and CN-Celeb verify the effectiveness of the three proposed architectures.
One of these architectures achieves nearly a 23\% lower Equal Error Rate compared to that of ECAPA-TDNN on VoxCeleb1-O dataset, demonstrating the competitive performance achievable among the current TDNN architectures under the comparable parameter count.
\end{abstract}

\begin{IEEEkeywords}
Speaker verification, ECAPA-TDNN, Res2Net, contextual information
\end{IEEEkeywords}

\IEEEpeerreviewmaketitle

\section{Introduction}
\IEEEPARstart{S}{peaker} verification (SV) is a key task in the field of speech deep learning, which aims to verify whether the test utterance and the registered utterance are from the same speaker based on their speech features.
SV can be divided into text-dependent SV and text-independent SV according to whether the content of the utterance spoken by the speaker is specified. In this paper, we focus on the text-independent SV task. 
A typical SV system consists of a speaker embedding encoder in the front end and a scoring back-end.
Traditional encoders and scoring back-ends used to be dominated by i-vector \cite{i-vector} systems, cosine similarity scoring \cite{cosine}, and probabilistic linear discriminant analysis (PLDA) \cite{PLDA}.
Recently, remarkable advancements have been achieved in the research of SV due to the continuous development of deep learning technology.
Since the introduction of x-vector \cite{x-vector}, deep learning has increasingly been applied to SV tasks, becoming the predominant approach \cite{x-vector-sr}. This trend has catalyzed a surge of research focused on speaker embedding extractors, utilizing the Time Delay Neural Network (TDNN) \cite{tdnn} architecture.
Additionally, the ResNet {module} \cite{resnet}, which is renowned for its excellent performance in the visual field, has been integrated as a speaker embedding encoder within SV frameworks.
The residual connections in ResNet can effectively avoid gradient vanishing and enhance model performance, particularly in substantially deep networks. 
However, just using convolutions {in ResNet} ignores the inter-dependencies between channels. To address this limitation, \cite{se_sv} combined the Squeeze-Excitation (SE) attention network \cite{se} with ResNet to enhance the ability of the model for speaker information representation. The emergence of ECAPA-TDNN \cite{ecapa} has elevated the performance of SV systems to a new level. This architecture integrates SE block and 1-dimensional (1D) Res2Net \cite{res2net} module {with dilated convolutions} to extract multi-scale features with channel attention, {termed SE-Res2Block}. Besides, it employs {attentive statistical pooling} to further aggregate features, thereby improving the overall model performance.

Although the existing TDNN-based architectures excel in modeling local information, their capabilities for global and long-term modeling remain slightly insufficient, and there is still a gap compared to 2-dimensional (2D) convolutions. To address this gap, Branch-ECAPA-TDNN \cite{yaobranch} has been proposed to construct two separate branches for local and global information modeling by employing both convolutions and self-attention mechanism. In \cite{DS-TDNN}, a TDNN module, termed Global-aware Filter layer, is proposed to effectively extract global features.
In \cite{pcf}, the proposed architecture called PCF-ECAPA splits the spectrogram into multiple frequency bands and fuses the bands so as to compensate for the deficiency of ECAPA-TDNN in capturing time-frequency relevance within the spectrogram, which is a capability inherent in 2D convolutional models.
In \cite{ecapa++}, an enhanced ECAPA-TDNN architecture, named as ECAPA++, is introduced, in which the recursive convolution (RecConv) is proposed to replace the original convolution in Res2Net to better capture fine-grained speaker information. Besides, it allows significant network depth increase while maintaining model complexity.
These methods all improve the global information modeling ability of ECAPA-TDNN and the performance of SV tasks to a certain degree.
{Nevertheless}, the Res2Block convolutional blocks in ECAPA-TDNN exclusively attend to the {preceding }information and completely disregards the {subsequent} information in modeling multi-scale features. To fully exploit contextual information, we propose three alternative blocks for SE-Res2Block in ECAPA-TDNN to extract multi-scale features capable of modeling context dependence, which are:
\begin{itemize}
	\item \textbf{SE-Bi-Res2Block}: a {bi-directional} Res2Block structure is used to model multi-scale contextual information;
	\item \textbf{Bi-SE-Res2Block}: a {bi-directional} SE-Res2Block is used to extract multi-scale contextual information, and at the same time, high-level features are further extracted and fused by increasing the depth of the network through which the forward and reverse information pass alone.
	\item \textbf{SE-Res2Bi-LSTM Block}: The {dilated} convolutions in {Res2Blocks} are replaced by Bi-LSTM modules, leveraging their powerful capability in modeling long short-term contexts to enhance model performance.
\end{itemize}




The remainder of the paper is organized as follows. Section \ref{Proposed-Architectures} describes the three alternative blocks proposed for SE-Res2Block. Section \ref{Experiments} explains the experimental setup and Section \ref{EXp-result} discusses the results of experiments. A brief conclusion is drawn in Section \ref{conclusion}.

\section{Proposed Blocks}
\label{Proposed-Architectures}

%

\subsection{SE-Bi-Res2Block}

\begin{figure}[!t]
 \centering
 \includegraphics[scale=0.9]{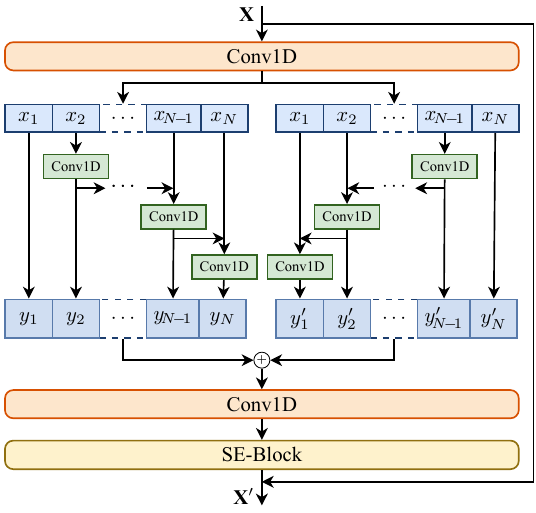}
 \caption{The proposed SE-Bi-Res2Block that consists of Res2Block and reversed Res2Block.}
 \label{SE-Bi-Res2Net}
\end{figure}

The one-way structure in the SE-Res2Block can only aggregate information from the {preceding} feature maps and ignore the impact of the {subsequent information} on the current feature map subset. To address this limitation, we design a {bi-directional} Res2Net structure as shown in Fig.\thinspace\ref{SE-Bi-Res2Net}, denoted as SE-Bi-Res2Block, which is inspired by the approach in \cite{PhaseDCN} to extract multi-scale information from two directions in speech enhancement. Our aim is to extract multi-scale contextual information contained in the entire feature map from two directions.

The {proposed SE-Bi-Res2Block} operates by initially feeding the feature map ${\mathbf X}$ through a {1D convolutional layer with kernel size $k=1$} to adjust the number of feature channels to a size which is divisible by the scale dimension $N$. The output feature map is then split into $N$ subsets. Two multi-scale features are extracted from the two directions before and after the current feature map subset. The two multi-scale features are then added, and the fused {bi-directional} multi-scale features are fed into {a 1D convolutional layer with kernel size $k=1$} to recover the number of channels. Finally, an SE-Block is used for channel attention processing. The whole process can be expressed by the following formula, 
\begin{equation}
	\mathbf{X^{\prime}}\! =\! {\mathrm{SE}} \left\{ f_{2} \left[ {\mathrm{Res2}}\left(f_{1}\left( \mathbf{X} \right) \right) + {\mathrm{Res2\_Rev}}\left(f_{1}\left( \mathbf{X} \right) \right) \right] \right\}+{\mathbf{X}}
\end{equation}
where $\mathbf{X^{\prime}}$ represents output feature map of the SE-Bi-Res2Block. $f_1$ and $f_2$ represent two 1D convolution {layers}, while
${\mathrm{Res2}}$ and ${\mathrm{Res2\_Rev}}$ signify {Res2Block and reversed Res2block} processing. Additionally, ${\mathrm{SE}}$ refers to a squeeze-and-excitation block.
${\mathrm{Res2}}$ splits the feature map $f_1(\mathbf{X})$ evenly into $N$ feature map subsets along the channel dimension, denoted as $\{x_1,x_2,\cdots,x_N\}$. The first feature map subset $x_1$ is directly used as the output $y_1$ without any processing, aiming at parameter reduction when the scale dimension $N$ increases.
Starting from $x_2$, each feature map subset $x_i$ is fed into a 1D convolutional layer with kernel size $k=3$, represented as $\mathbf{K}_i$. The convolved output $y_i$ is then added to $x_{i+1}$ and fed into the next 1D convolutional layer $\mathbf{K}_{i+1}$, $i=2,\cdots,N-1$. This process can be formulated as,
\begin{equation}
	y_{i}=\left\{\begin{array}{ll}
		\!\!x_{i}, & i=1 \\
		\!\!\mathbf{K}_{i}\left(x_{i}\right), & i=2 \\
		\!\!\mathbf{K}_{i}\left(x_{i}+y_{i-1}\right), & i=3,4, \cdots, N
	\end{array}\right.
\end{equation}
Similarly, the process of ${\mathrm{Res2\_Rev}}(f_1({\mathbf X}))$ can be formulated as,
\begin{equation}
	y_{i}^{\prime}=\left\{\begin{array}{ll}
		\!\!x_{i}, & i=N \\
		\!\!\mathbf{K}_{i}\left(x_{i}\right), & i=N-1 \\
		\!\!\mathbf{K}_{i}\left(x_{i}+y^{\prime}_{i+1}\right), & i=N-2,\cdots,2,1
	\end{array}\right.
\end{equation}
Finally, the outputs $\{y_1,y_2,\cdots,y_N\}$ and $\{y_1^{\prime},y_2^{\prime},\cdots,y_N^{\prime}\}$ corresponding to the feature map subsets are {added} and reassembled into a new feature map.

\subsection{Bi-SE-Res2Block}

\begin{figure}[!t]
 \centering
 \includegraphics[scale=0.9]{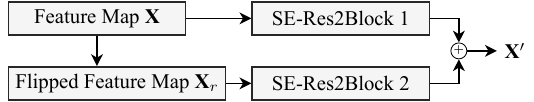}
 \caption{The proposed Bi-SE-Res2Block.}
 \label{Bi-SE-Res2Net}
\end{figure}

Considering the potential limitation of one additional reversed Res2Block in effectively extracting comprehensive multi-scale features that fully exploit the contextual information, we design a {dual-stream module}, depicted in Fig.\thinspace\ref{Bi-SE-Res2Net}, termed Bi-SE-Res2Block.
Prior to entering the SE-Res2Block, the feature map ${\mathbf X}$ undergoes a channel-wise flip to obtain ${\mathbf X}_r$. Subsequently, ${\mathbf X}$ is processed through an SE-Res2Block to extract multi-scale features utilizing preceding information, while ${\mathbf X}_r$ is fed into another SE-Res2Block to capture multi-scale features using the subsequent information. Finally, the multi-scale contextual information ${\mathbf X}^{\prime}$ is obtained by aggregating the multi-scale features from the two streams.
The whole process can be expressed as, 
\begin{equation}
	\mathbf{X^{\prime}} = {\mathrm{SE\!-\!Res2Block}}_1\left( \mathbf{X} \right) +{\mathrm{SE\!-\!Res2Block}}_2\left( \mathbf{X}_r \right)
\end{equation}

\subsection{SE-Res2Bi-LSTM Block}
\label{Res2Bi-LSTM}
As previously mentioned, Res2Block generates richer and more diverse features by aggregating unique parallel branches to capture multi-scale information.
However, the model treats all the information contained in each feature map subset as equally important in the subsequent convolution and feature aggregation process, which weakens the relatively top feature map subset and makes the model unable to focus on the key information in the context.
The model also cannot distinguish between relevant and irrelevant information in the context.
As a result, while Res2Block demonstrates effectiveness in establishing short-term dependencies for modeling long time sequence such as speeches, its capability to establish long-term dependencies remains somewhat inadequate.
Consequently, despite employing bi-directional Res2Block or bi-directional SE-Res2Block architecture, the challenge of insufficient capture of long-term dependencies still persists.

To tackle this challenge, we substitute the central convolutional layers in SE-Bi-Res2Block with LSTM modules and integrate hierarchical residual connections to manage multi-scale features. This approach is intended to utilize the long and short-term dependencies inherent in LSTM for extracting multi-scale information from feature map subsets. {Specifically, it focuses on extracting relevant information while disregarding irrelevant details. However, it is evident that the extracted feature maps display information redundancy. Moreover, the introduction of bi-directional structure has notably increased the complexity of the modified block.}

\begin{figure}[!t]
\centering
\includegraphics[scale=0.9]{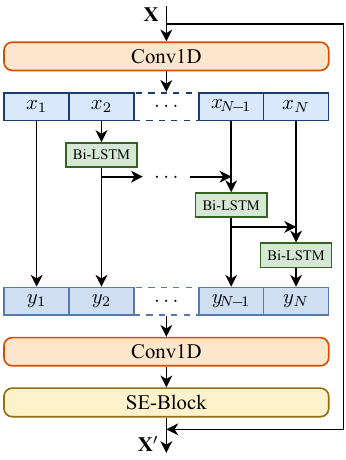}
\caption{The proposed SE-Res2Bi-LSTM block.}
\label{SE-Res2Bi-LSTM}
\end{figure}

For the purpose of establishing short and long-term dependencies as well as restricting model complexity, we replace the central convolutional layers of Res2Block with Bi-LSTM modules, introducing the SE-Res2Bi-LSTM block, as illustrated in Fig.\thinspace\ref{SE-Res2Bi-LSTM}.
This approach exploits the robust long-short term dependency characteristics inherent in Bi-LSTM to model context effectively. Simultaneously, the SE-Res2Bi-LSTM model mitigates the interference of irrelevant information and minimizes overall feature redundancy without overly escalating model complexity.

\section{Experimental setup}
\label{Experiments}
\subsection{Datasets}

Various datasets are utilized in the experiments, including VoxCeleb1 \cite{voxceleb1_2017, voxceleb1_2020}, VoxCeleb2 \cite{voxceleb2}, CN-Celeb \cite{cnceleb1}, MUSAN \cite{musan}, and RIR \cite{rir}.
Architectures incorporating the proposed three blocks are evaluated and two sets of experiments are implemented. For the first set, the training set comprises the development set of VoxCeleb2, while VoxCeleb1-O, VoxCeleb1-E, and VoxCeleb1-H are employed as test sets to evaluate the performance of the proposed architectures. {The second set evolves the large-scale CN-Celeb dataset which covers much more genres of speeches than VoxCeleb.} The development set from CN-Celeb.T is utilized as training set, with evaluations performed on the test set CN-Celeb.E.
The CN-Celeb.T training set {utilized in our experiments} comprises 2,796 speakers and 632,740 {utterances}, whereas the CN-Celeb.E test set includes 200 speakers and 17,973 {utterances}. We perform speaker validation through cross-pairing between registered and test speeches, yielding 3,484,292 test pairs that are independent of gender.
Furthermore, the MUSAN and RIR datasets are employed for online augmentation, introducing noise and reverberation, respectively.

{To verify the effectiveness of the proposed architectures on complicated multi-scale datasets, }different types of training experiments are conducted, including single-genre training (SG training), multi-genre training (MG training), and mixed training (M training) {\cite{cnceleb2}.} {Specifically, the development set in VoxCeleb2 is used for SG training, the development set in CN-Celeb.T for MG training, and a combination of development sets in VoxCeleb2 and CN-Celeb.T for M training.}

\subsection{Experimental setup}
\label{exp_set}
For all the experiments, {the audio input duration is {standardized} into 2 seconds. During data augmentation, the MUSAN dataset and RIR dataset are used for online noise addition and simulated reverberation, respectively. The model takes an 80-dimensional Mel spectral feature as input, which is extracted using a {Hamming window} of size 25ms and a frame shift of 10ms. Additionally, specaugment technique \cite{specaugment} is applied for time-frequency masking, with maximum masking band size set to 8 for frequency domain and 10 for time domain, respectively.}


The loss function employed is the additive angular margin softmax (AAM-Softmax) loss \cite{arcface, margin}, with the loss margin and loss scale set to 0.2 and 30, respectively. Throughout training, the Adam optimizer \cite{adam} is utilized with weight decay set to 2e-5. {The cyclical learning rate scheduler is adopted with the triangular2 policy \cite{cyclical}, where the learning rate varies between 1e-8 and 1e-3, and undergoes 65k iterations of up and down steps.}
The mini-batch size during training is consistently set to 128.
Adaptive score normalization (AS-norm) is used for score normalization and
the evaluation metrics are Equal Error Rate (EER) and the Minimum Detection Cost Function (MinDCF), with hyper-parameters set as $P_{\mathrm{target}} = 0.01$ and $C_{\mathrm{FA}} = C_{\mathrm{Miss}} = 1$.
The Architectures incorporating the proposed three blocks are designated as SE-Bi-Res2Block-ECAPA, Bi-SE-Res2Block-ECAPA, and SE-Res2Bi-LSTM-ECAPA, respectively.
The number of filters $C$ in the convolutional layers, also referred to as system width, is set to either 512 or 1024\footnote{The source code is available at https://github.com/wsdragon2010/Res2Bi-LSTM}.

\begin{table*}[t] 
	\centering
	\caption{EER and MinDCF performances of various systems with different width on VoxCeleb1-O, VoxCeleb1-E and VoxCeleb1-H.}
	\label{t1}
	\noindent
	\renewcommand{\arraystretch}{1.29} 
	\begin{tabular}{m{4.3cm}<{\centering} m{1.3cm}<{\centering}m{1cm}<{\centering}m{1cm}<{\centering}m{0.05cm}<{\centering}m{1cm}<{\centering} m{1cm}<{\centering}m{0.05cm}<{\centering}m{1cm}<{\centering}m{1cm}<{\centering}<{\centering}}
		\toprule 
		\multirow{2}{*}{\bf Model} & \multirow{2}{*}{\bf \#Para.} & \multicolumn{2}{c }{\bf VoxCeleb1-O} & \multicolumn{1}{c }{} & \multicolumn{2}{c }{\bf VoxCeleb1-E} & \multicolumn{1}{c }{} & \multicolumn{2}{c }{\bf VoxCeleb1-H} \\ 
		\cline{3-4} \cline{6-7} \cline{9-10} 
		& & \bf EER(\%) & \bf MinDCF & & \bf EER(\%) & \bf MinDCF & & \bf EER(\%) & \bf MinDCF \\ 
		\cline{1-10} 
		ECAPA-TDNN (C=512) (2019) & 6.2M& 1.01 & 0.1274 & & 1.24 & 0.1418 & & 2.32 & 0.2181 \\
		ECAPA-TDNN (C=1024) (2019) & 14.73M& 0.87 & 0.1066 & & 1.12 & 0.1318 & & 2.12 & 0.2101\\
		\cline{1-10} 
		ResNet34-DTCF (2021) \cite{DTCF}& $\approxeq$ 9M & 0.79 & 0.1090 & & 1.13 & 0.1259 & & 2.09 & 0.2082 \\
		Branch-ECAPA(2023) \cite{yaobranch} & 24.11M & 0.718 & \textbf{0.084} & & 0.916 & 0.098 & & 1.69 & 0.166 \\
		PCF-ECAPA (2023) \cite{pcf}& 22.2M & 0.718 & 0.892 & & 0.891 & 0.1024 & & 1.707 & 0.1754 \\
		ECAPA++ (2023) \cite{ecapa++}& 14.7M & 0.76 & 0.096 & & \textbf{0.84} & \textbf{0.0981} & & \textbf{1.54} & \textbf{0.1536} \\
		DS-TDNN-B (2024) \cite{DS-TDNN}& 13.2M & 0.78 & 0.092 & & 1.06 & 0.126 & & 1.86 & 0.186 \\
		\cline{1-10} 
		SE-Bi-Res2Block-ECAPA (C=512) & 6.90M & 0.93 & 0.1459 & & 1.22 & 0.1281 & & 2.22 & 0.2606 \\
		SE-Bi-Res2Block-ECAPA (C=1024) & 15.72M & 0.81 & 0.1394 & & 1.05 & 0.1120 & & 1.97 & 0.2470 \\
		Bi-SE-Res2Block-ECAPA (C=512) & 8.79M & 0.85 & 0.1355 & & 1.11 & 0.1145 & & 2.00 & 0.2362\\
		Bi-SE-Res2Block-ECAPA (C=1024) & 22.49M & 0.75 & 0.1083 & & 1.02 & 0.1103 & & 1.90 & 0.2396\\
		{SE-Res2Bi-LSTM-ECAPA} (C=512) & 6.91M & 0.83 & 0.1332 & & 1.11 & 0.1179 & & 2.01 & 0.2324 \\
		{SE-Res2Bi-LSTM-ECAPA} (C=1024) & 15.73M & \textbf{0.67} & 0.1108 & & 0.99 & 0.1069 & & 1.82 & 0.2212 \\
		\bottomrule 
	\end{tabular}
\end{table*}

\begin{table}[t] 
	\centering
	\noindent
 	\renewcommand{\arraystretch}{1.29} 
 	\setlength{\tabcolsep}{2.5pt} 
	\caption{Performance of the proposed architectures with $C=1024$ under different training types on CN-Celeb.E.}
	\label{t2}
	\begin{tabular}{m{3.85cm}<{\centering}m{1.3cm}<{\centering}m{1cm}<{\centering}m{1cm}<{\centering}}
		\toprule 
		\multirow{2}{*}{\bf Model} & \multirow{2}{*}{\bf \#Para.} & \multicolumn{2}{c }{\bf CN-Celeb.E}\\ 
		\cline{3-4}
		& & \bf EER(\%) & \bf MinDCF\\ 
		\cline{1-4} 
		ResNet34-DTCF (2021) \cite{DTCF}& $\approxeq$ 9M & 14.84 & 0.5961 \\
		Branch-ECAPA (M) (2023) \cite{yaobranch}& 25.71M & 6.922 & 0.357 \\
		CAM++ (M) (2023) \cite{cam++}& 7.18M & 6.78 & 0.3830 \\
		\cline{1-4} 
		SE-Bi-Res2Block-ECAPA (SG) & 15.72M & 13.44 & 0.4513\\
		SE-Bi-Res2Block-ECAPA (MG) & 15.72M & 8.36 & 0.4319\\
		SE-Bi-Res2Block-ECAPA (M) & 15.72M & 6.89 & 0.3639 \\
		
		\cline{1-4} 
		Bi-SE-Res2Block-ECAPA (SG) & 22.49M & 12.73 & 0.4324 \\
		Bi-SE-Res2Block-ECAPA (MG) & 22.49M & 7.95 &0.4284 \\
		Bi-SE-Res2Block-ECAPA (M) & 22.49M & \textbf{6.54} & \textbf{0.3517} \\
		
		\cline{1-4} 
		{SE-Res2Bi-LSTM-ECAPA} (SG) & 15.73M & 12.44 & 0.4335 \\
		{SE-Res2Bi-LSTM-ECAPA} (MG) & 15.73M & 8.13 & 0.4316 \\
		{SE-Res2Bi-LSTM-ECAPA} (M) & 15.73M & 6.63 & 0.3565 \\
		\bottomrule 
	\end{tabular}
\end{table}

\section{Experimental results}
\label{EXp-result}

The EER and minDCF performances of ECAPA-TDNN, five State-of-the-Art (SOTA) TDNN-based SV systems, and the proposed architectures on VoxCeleb1 are presented in Table \ref{t1}.
The EER of the three proposed architectures outperforms that of ECAPA-TDNN with the same system width on the three test sets, and {SE-Res2Bi-LSTM-ECAPA (C=1024)} exhibits the best performance, achieving EER improvements of 23\%, 11.6\%, and 14.2\% over ECAPA-TDNN on VoxCeleb1-O, VoxCeleb1-E and VoxCeleb1-H, respectively, while the number of parameters in SE-Res2Bi-LSTM-ECAPA (C=1024) is only {1 million} higher than that of ECAPA-TDNN.
Additionally, the EER of SE-Res2Bi-LSTM-ECAPA (C=1024) on VoxCeleb1-O surpasses all the five SOTA TDNN-based SV systems, even though two of them have greater parameter counts.
Among the three proposed architectures, {SE-Res2Bi-LSTM-ECAPA} performs best, followed by Bi-SE-Res2Block-ECAPA with the highest number of parameters, and the worst is SE-Bi-Res2Block-ECAPA. This result is also in line with our theoretical expectations as mentioned earlier.
However, the performance of the proposed three structures decreases in terms of MinDCF on VoxCeleb1-O and VoxCeleb1-H, while showing improvement on VoxCeleb1-E to some extent.

Table \ref{t2} shows the EER and minDCF performances of the proposed architectures with system width $C=1024$ on CN-Celeb.E. Besides, different training types are considered. It is evident that {EER and MinDCF of Bi-SE-Res2Block-ECAPA are the best under M training.} This underscores the effectiveness of the Bi-SE-Res2Block in capturing deep contextual multi-scale features which are essential for tasks with higher complexity. 

\section{Conclusion}
\label{conclusion}
In this paper, three enhanced architectures based on ECAPA-TDNN for the speaker verification task are proposed. The goal is to address the limitation of Res2Block in effectively modeling long-term contextual dependencies. The three architectures make up for this deficiency layer by layer through the use of {bi-directional} Res2Blocks, {bi-directional} SE-Res2Blocks and Bi-LSTM to enhance the modeling of long and short-term contextual dependencies.
The effectiveness of these architectures is verified through experimental performances on VoxCeleb and CN-Celeb. Notably, the EER of {SE-Res2Bi-LSTM-ECAPA} achieves 23\%, 11.6\% and 14.2\% performance improvement on VoxCeleb1-O, E and H, respectively, with only 1 million additional parameters comparing to ECAPA-TDNN.

\bibliographystyle{IEEEtran}
\bibliography{IEEEabrv, refs.bib}

\end{document}